\documentclass[pdftex,twocolumn,epjc3]{svjour3}          
\RequirePackage{flushend}
\RequirePackage[numbers,sort&compress]{natbib}
\RequirePackage[colorlinks,citecolor=blue,urlcolor=blue,linkcolor=blue]{hyperref}
\usepackage{amsmath,lipsum,amssymb}
\usepackage{cuted}
\usepackage{latexsym}
\usepackage{amsfonts}
\usepackage{graphicx}
\usepackage{mathrsfs}
\usepackage{epstopdf}
\usepackage{bm}
\usepackage{orcidlink}

\journalname{Eur. Phys. J. C}
\begin{document}
\title{Inspiraling binary charged black holes in an external magnetic field: Application of post-Newtonian dynamics in Einstein-Maxwell theory}
\author{RunDong Tang  \thanksref{e1,addr1,addr2} \and  Lang Liu \thanksref{e2,addr3} \and Wen-Biao Han\thanksref{e3,addr1,addr2,addr4,addr5,addr6}}
\thankstext{e1}{e-mail: trd@shao.ac.cn}
\thankstext{e2}{e-mail: liulang@bnu.edu.cn  (corresponding author)}
\thankstext{e3}{e-mail: wbhan@shao.ac.cn  (corresponding author)}

\institute{Shanghai Astronomical Observatory, Chinese Academy of Sciences, 80 Nandan Rd, Shanghai 200030, China \label{addr1} \and School of Astronomy and Space Science, University of Chinese Academy of Sciences, 19 Yuquan Rd, Beijing 100049, China \label{addr2} \and Department of Physics, Faculty of Arts and Sciences, Beijing Normal University, 18 Jinfeng Rd, Zhuhai 519087, China \label{addr3} \and School of Fundamental Physics and Mathematical Sciences, Hangzhou Institute for Advanced Study, UCAS, Hangzhou 310024, China \label{addr4} \and Taiji Laboratory for Gravitational Wave Universe (Beijing/Hangzhou), University of Chinese Academy of Sciences, Beijing 100049, China \label{addr5} \and Key Laboratory of Radio Astronomy and Technology, Chinese Academy of Sciences, A20 Datun Road, Chaoyang District, Beijing 100101, China \label{addr6}}
\date{Received: date / Accepted: date}

\maketitle

\begin{abstract}
We present a systematic post-Newtonian treatment of binary charged black holes immersed in external magnetic fields within the framework of Einstein-Maxwell theory. By incorporating a uniform external magnetic field into the two-body Lagrangian expanded to first post-Newtonian order, we derive the complete equations of motion that capture both gravitational and electromagnetic interactions. The magnetic Lorentz force fundamentally alters the orbital dynamics, breaking the conservation of linear and angular momentum and inducing transitions from planar to three-dimensional trajectories. {Through numerical integration of these equations, we compute the resulting gravitational waveforms and characterize the distinctive magnetic field signatures through time-domain and frequency-domain analysis.} Our results demonstrate that strong background magnetic fields can substantially modify the orbital evolution and leave distinctive signatures in the gravitational wave signals. These findings provide a promising avenue for detecting charged black holes and probing magnetic field environments through gravitational wave observations.
\end{abstract}

\section{Introduction}
The detection of charged black holes represents a fundamental frontier in gravitational wave astronomy and tests of general relativity. The groundbreaking observations by the LIGO-Virgo-KAGRA Collaboration \cite{Abbott:2016blz} have opened unprecedented opportunities for probing the nature of black holes through gravitational wave astronomy \cite{LIGOScientific:2018mvr, LIGOScientific:2020ibl, LIGOScientific:2021djp,LIGOScientific:2025slb}, potentially enabling searches for deviations from the Kerr paradigm \cite{Yunes:2016jcc,Barack:2018yly}. According to the no-hair theorem, astrophysical black holes are uniquely characterized by three parameters: mass, angular momentum, and electric charge \cite{Israel:1967wq,Carter:1971zc,Robinson:1975bv}. While the first two parameters are routinely measured in gravitational wave observations, the potential existence of charge remains an open question with profound implications for fundamental physics \cite{Cardoso:2016ryw}.

Binary systems containing charged black holes exhibit fundamentally richer dynamics than their neutral counterparts. In comparison with Schwarzschild black holes, charged black holes emit not only gravitational radiation but also electromagnetic radiation, leading to qualitatively different phenomena and opening new observational channels \cite{Zilhao:2012gp,Zilhao:2013nda,Liebling:2016orx,Toshmatov:2018tyo,Zhang:2023tfv,Pereniguez:2023wxf,Bhattacharyya:2023kbh,Zhang:2023zmb,DeFelice:2023rra,Pereniguez:2024fkn,Grilli:2024fds,Ghosh:2025ban,Pereniguez:2025jxq}. This dual emission mechanism has sparked increasing theoretical interest in recent years, as evidenced by extensive investigations into charged black hole physics and their observational signatures (see Refs.~\cite{Maldacena:2020skw,Bai:2020spd,Liu:2020vsy,Ghosh:2020tdu,Liu:2020bag,Bai:2019zcd,Liu:2020cds,Allahyari:2019jqz,Christiansen:2020pnv,Wang:2020fra,Bozzola:2020mjx,Kim:2020bhg,Cardoso:2020nst,McInnes:2020gxx,Bai:2020ezy,Diamond:2021scl,Bozzola:2021elc,McInnes:2021frb,Kritos:2021nsf,Hou:2021suj,Benavides-Gallego:2021the,Diamond:2021dth,Ackerman:2008kmp,Feng:2009mn,Foot:2014uba,Foot:2014osa,Moffat:2005si,Cardoso:2016olt,Cardoso:2020iji,Liu:2022wtq,Zhang:2022hbt,Pina:2022dye,Benavides-Gallego:2022dpn,Liu:2022xtn,Chen:2022qvg,Estes:2022buj} and references therein). The electromagnetic interactions introduce additional energy and angular momentum loss channels \cite{Liu:2020cds,Christiansen:2020pnv}, modify the gravitational wave phase evolution \cite{Bozzola:2020mjx,Bozzola:2021elc}, and potentially enable multi-messenger observations that combine gravitational and electromagnetic signals \cite{Liebling:2016orx,Cardoso:2016ryw}. These distinctive features make charged black holes particularly interesting targets for next-generation gravitational wave detectors, as the deviations from pure gravitational radiation could provide direct evidence for black hole charge. Recent theoretical developments have established the post-Newtonian (PN) framework for charged compact objects \cite{Khalil:2018aaj,Gupta:2022spq, Henry:2023guc,Placidi:2025xyi}, providing the foundation for precision modeling of such systems and enabling quantitative predictions for their observational signatures.

A crucial yet underexplored aspect is the influence of external magnetic fields on charged black hole dynamics. Astrophysical environments are permeated by magnetic fields spanning an enormous range of scales and strengths \cite{Kronberg:1993vk,Durrer:2013pga,Subramanian:2015lua}. From the $\sim 10^{-6}$ G fields in galactic halos \cite{Beck:2015jta} to the $\sim 10^{15}$ G fields near magnetars \cite{Kaspi:2017fwg}, these fields are ubiquitous features of the cosmic landscape. For charged black holes moving through such magnetized environments, the Lorentz force fundamentally alters the orbital dynamics, breaking the symmetries that govern neutral binary evolution.

Accurate gravitational wave template construction requires incorporating all relevant physical effects to the appropriate PN order. The PN formalism, based on systematic expansions in the characteristic velocity $v/c$ and gravitational potential $GM/rc^2$, has been extensively developed for neutral binaries \cite{Blanchet:2013haa,Thorne:1980ru}. {Similar post-Newtonian techniques have been successfully applied to study compact binary dynamics in modified gravity theories, including recent systematic developments in Einstein-Cartan theory \cite{Battista:2022sci, Battista:2023znv,DeFalco:2024ojf}.} For charged systems, recent works have extended this framework to include electromagnetic interactions \cite{Henry:2023guc,Placidi:2025xyi}. However, the coupling to external electromagnetic fields-a generic feature of realistic astrophysical environments remains largely unexplored within the PN framework. This gap is particularly significant given that external fields can dominate over binary electromagnetic interactions in many astrophysical scenarios.

In this work, we develop a PN treatment of binary charged black holes immersed in external magnetic fields. By incorporating the magnetic field interaction into the Einstein-Maxwell Lagrangian at first PN order, we derive the complete equations of motion accounting for both gravitational and electromagnetic effects. We demonstrate that external magnetic fields fundamentally alter the conservation laws, induce three-dimensional orbital motion, and leave distinctive signatures in the gravitational waveforms. Our analysis provides the theoretical foundation for using gravitational wave observations to probe both the charge of black holes and the magnetic field environments in which they evolve. The paper is organized as follows. In Section \ref{sec:action}, we construct the PN Lagrangian for a spinless two-body system in Einstein-Maxwell theory (EMT), incorporating the interaction with an external uniform magnetic field. Section \ref{sec:solution} presents the solutions to the equations of motion, analyzing the evolution of energy, linear momentum, and angular momentum, with detailed comparisons between magnetized and non-magnetized systems. In Section \ref{sec:waveform}, we compute the gravitational waveforms and quantify the magnetic field imprints through matched filtering analysis. Finally, Section \ref{sec:conclusion} summarizes our findings and discusses their implications for gravitational wave astronomy and multi-messenger observations.

\section{Post-Newtonian expansion in Einstein-Maxwell theory}\label{sec:action}
\subsection{Action in Einstein-Maxwell theory}

We adopt the EMT framework and employ the PN formalism, extending the standard approach used in pure gravitational systems. A crucial distinction arises when transitioning from flat to curved spacetime: while electromagnetic dynamics remains linear in Minkowski space—with the electromagnetic field not serving as a source for Maxwell's equations—the situation changes fundamentally in EMT. The coupling between gravitational and electromagnetic fields renders electrodynamics inherently nonlinear in curved spacetime, necessitating a careful treatment of field interactions.

For inspiraling black hole binaries, the dynamics can be systematically analyzed through a formal Taylor expansion in the PN parameter $\epsilon \sim v/c \sim (Gm/rc^2)^{1/2}$, where $v$ represents the characteristic orbital velocity, $m$ the typical mass scale, and $r$ the orbital separation. At leading order, the dynamics reduces to the Newtonian limit with successive relativistic corrections appearing at higher orders. Throughout this work, we employ Gaussian units and retain the fundamental constants $G$ and $c$ explicitly to track the PN order systematically. Terms of order $\mathcal{O}((v/c)^{2n})$ are designated as $n$PN corrections relative to the Newtonian dynamics.

The complete treatment would require solving the coupled system of equations of motion for the sources and field equations, followed by the implementation of the point-particle approximation and regularization of divergences associated with Dirac delta functions, as established in previous foundational works \cite{Blanchet:2013haa,Blanchet:1998vx}. For our present purposes, however, it proves more efficient to directly employ the matter action for an $N$-body system (with $N=2$ for binary systems) in EMT, following the formulation developed in recent studies \cite{Henry:2023guc}
\begin{equation}
    S_\text{m} = \sum_{a=1}^{N} \left[ -m_a c \int \sqrt{-g_{\mu\nu} v_a^\mu v_a^\nu} \, dt + \frac{q_a}{c} \int v_a^\mu A_\mu \, dt \right],    
\end{equation}
where particle $a$ is characterized by mass $m_a$, electric charge $q_a$, and coordinate velocity $v_a^\mu = dx_a^\mu/dt$ with $x_a^\mu$ denoting its spacetime coordinates. The first integral represents the standard geodesic action in curved spacetime, while the second describes the minimal coupling between charged particles and the electromagnetic four-potential $A_\mu$. 

The field sector action, encompassing both gravitational and electromagnetic contributions, takes the standard form \cite{Landau:1975pou}
\begin{equation}
    S_\text{F} = \int d^4x \sqrt{-g} \left[ -\frac{c^3}{16\pi G} R - \frac{1}{16\pi c} F^{\mu\nu} F_{\mu\nu} \right],
\end{equation}
where $R$ denotes the Ricci scalar curvature, $F_{\mu\nu} = \nabla_\mu A_\nu - \nabla_\nu A_\mu = \partial_\mu A_\nu - \partial_\nu A_\mu$ is the electromagnetic field tensor, and $g = \det(g_{\mu\nu})$. The volume element is expressed as $d^4x = c \, dt \, d^3x$. While we adopt the standard Einstein-Hilbert form for the gravitational action, we note that the alternative Landau-Lifschitz formulation \cite{Landau:1975pou}, containing only first-order derivatives of the metric, differs merely by a total derivative and yields equivalent equations of motion.  

The electromagnetic field tensor relates to the physical electric and magnetic fields through
\begin{equation}
    E^i = F^{0i}, \quad B^i = \frac{1}{2} \epsilon^{ijk} F_{jk},
\end{equation}
where $\epsilon^{ijk}$ denotes the three-dimensional Levi-Civita symbol. Throughout this work, we adopt the standard convention where Greek indices $\mu, \nu \in \{0,1,2,3\}$ span spacetime coordinates, Latin indices $i,j,k \in \{1,2,3\}$ denote spatial components, and all indices are raised and lowered using the metric tensor $g_{\mu\nu}$.

The complete action governing the coupled Einstein-Maxwell system is
\begin{equation}\label{eq:total action}
    S = S_\text{m} + S_\text{F}.
\end{equation}
And the gravitational-electromagnetic coupling manifests through the contraction $F^{\mu\nu} F_{\mu\nu} = g^{\mu\rho} g^{\nu\sigma} F_{\rho\sigma} F_{\mu\nu}$, where the metric tensor mediates the interaction between the two field sectors. This coupling becomes particularly significant in strong-field regimes and introduces qualitatively new effects absent in either pure gravitational or pure electromagnetic systems.

\subsection{External magnetic field}
Our investigation centers on binary charged black holes embedded in an external magnetic field, necessitating a careful treatment of the background electromagnetic environment. We decompose the total electromagnetic field tensor $\bar{F}_{\mu\nu}$ into contributions from the binary system and the external field
\begin{equation}\label{eq:total Maxwell equaiton}
  \bar{F}_{\mu \nu}=F_{\mu \nu}+\tilde{F}_{\mu \nu},
\end{equation}
where $F_{\mu\nu}$ represents the electromagnetic field generated by the charged black holes themselves, while $\tilde{F}_{\mu\nu}$ denotes the externally imposed magnetic field.

A key simplification in our treatment concerns the gravitational effects of the external magnetic field. We adopt the approximation that the background magnetic field does not contribute to the stress-energy tensor appearing in Einstein's field equations, thereby leaving the spacetime geometry unaffected by its presence. This approximation is well-justified even in extreme astrophysical environments: calculations for neutron stars, which possess the strongest known magnetic fields among compact objects, demonstrate that magnetic contributions to spacetime curvature remain negligible compared to matter contributions \cite{Aliev:1989wz,magnetic_field_of_neutron_star}.

{To quantify the validity of neglecting the gravitational backreaction of the magnetic field, we compare the magnetic energy density $u_B = B^2/8\pi$ to the characteristic gravitational energy density $\rho_{\rm grav} c^2 \sim M c^2/(\pi r^3)$. The ratio satisfies
\begin{equation}
\frac{u_B}{\rho_{\rm grav} c^2} \sim \frac{B^2 r^3}{M c^2} \sim \left(\frac{B}{B_{\rm crit}}\right)^2,
\end{equation}
where the critical field is $B_{\rm crit} \equiv \sqrt{8 M c^2/r^3}$. For stellar-mass binaries ($M \sim 35 M_\odot$) at separations $r \sim 100 , GM/c^2$, we find $B_{\rm crit} \sim 10^{15}$~G. The magnetic field strength used in our simulations ($B = 8.0 \times 10^{12}$~G) yields $u_B/\rho_{\rm grav} c^2 \sim 10^{-4}$, confirming that our approximation remains valid for $B \lesssim 10^{14}$~G across the parameter space considered.}

We further assume that the external magnetic field $\tilde{\boldsymbol{B}}$ maintains uniform strength and stationary configuration throughout the region occupied by the binary system. Mathematically, this translates to the condition
\begin{equation}
    \partial_\mu \tilde{B}_\nu = 0,
\end{equation}
implying that all spatial and temporal derivatives of the background field components vanish.

This decomposition establishes a clear hierarchy in our treatment: the internal electromagnetic field $F_{\mu\nu}$, generated by the orbital motion and intrinsic properties of the charged black holes, governs the electromagnetic interactions within the binary system. Meanwhile, the external field $\tilde{F}_{\mu\nu}$ breaks the isolation of the system and introduces additional electromagnetic forces that can significantly affect the orbital dynamics. Crucially, when iteratively solving the coupled Einstein-Maxwell equations, the stationary and uniform nature of $\tilde{\boldsymbol{B}}$ allows us to treat it as a fixed background that does not require dynamical evolution, considerably simplifying the computational framework.

\subsection{Equations of motion at the first post-Newtonian order}

The derivation of equations of motion for the binary system proceeds through a systematic PN expansion of the action (\ref{eq:total action}), valid in the weak-field, slow-motion regime characteristic of wide binary orbits. We implement the Fokker action method \cite{Fokker1929}, a powerful technique that eliminates field degrees of freedom by substituting regularized solutions of the field equations back into the action, yielding an effective action depending solely on the worldline variables of the constituent bodies.

The computational framework requires appropriate gauge choices to eliminate unphysical degrees of freedom and simplify the field equations. For the gravitational sector, we adopt the harmonic gauge condition
\begin{equation}
    g^{\rho\sigma} \Gamma^\mu_{\rho\sigma} = 0,
\end{equation}
where $\Gamma^\mu_{\rho\sigma}$ denotes the Christoffel symbols. This choice linearizes the principal part of Einstein's equations and facilitates the iterative solution procedure. {Correspondingly, for the electromagnetic sector, we impose the Lorenz gauge condition
\begin{equation}
    \nabla_\mu A^\mu \approx \partial_\mu A^\mu = 0,  
\end{equation}
where the approximation holds at leading post-Newtonian order since $\sqrt{-g} = 1 + \mathcal{O}(c^{-2})$ in our perturbative expansion around flat spacetime. } These gauge choices ensure consistency between the gravitational and electromagnetic sectors while maintaining the manifest covariance of the formulation.

For an isolated binary system without external fields, the 1PN-corrected Lagrangian for charged black holes takes the form
\begin{equation}\label{eq:Lagrangian for charged particles}
    L_{\rm{iso}}=-m_{1}c^2-m_{2}c^2+L_0+\frac{1}{c^2}L_1+\mathcal{O}(c^{-4}),
\end{equation}
where the Newtonian and 1PN contributions are given by
\begin{equation}
\begin{aligned}
    L_0 & =\frac{1}{2}m_{1}v_{1}^{2}+\frac{1}{2}m_{2}v_{2}^{2}+\frac{Gm_{1}m_{2}}{r}-\frac{q_{1}q_{2}}{r},\\
    L_1 & =\frac{1}{8}m_{1}v_{1}^{4}+\frac{1}{8}m_{2}v_{2}^{4}\\
    & +\frac{Gm_{1}m_{2}}{2r}\left[3(v_{1}^{2}+v_{2}^{2})-7\bm{v_{1}}\cdot \bm{v_{2}}-(\bm{n}\cdot \bm{v_{1}})(\bm{n}\cdot \bm{v_{2}})\right]\\
    & +\frac{q_{1}q_{2}}{2r}\left[\bm{v_{1}}\cdot \bm{v_{2}}+(\bm{n}\cdot \bm{v_{1}})(\bm{n}\cdot \bm{v_{2}})\right]\\
    & -\frac{G^{2}m_{1}m_{2}}{2r^2}(m_{1}+m_{2})-\frac{Gq_{1}q_{2}}{r^2}(m_{1}+m_{2})\\
    & -\frac{G}{2r^2}(m_{1}q_{2}^2+m_{2}q_{1}^2).
\end{aligned}
\end{equation}
In the neutral limit ($q_1 = q_2 = 0$), this expression reduces to the well-established Einstein-Infeld-Hoffmann (EIH) Lagrangian \cite{EIH-Lagrangian,Patil:2020dme,Khalil:2018aaj}, as verified in the harmonic-coordinate formulation by Blanchet and Damour \cite{deAndrade:2000gf}. The electromagnetic contributions preserve the fundamental conservation laws: the Lagrangian remains invariant under spatial rotations and spacetime translations, ensuring conservation of total energy, momentum, and angular momentum for the isolated system.

The introduction of an external magnetic field fundamentally alters this picture. For a uniform and stationary background field, we adopt the symmetric gauge representation
\begin{equation}
    \tilde{A}_0 = 0, 
 \quad   \tilde{\boldsymbol{A}} = \frac{1}{2} \tilde{\boldsymbol{B}} \times \tilde{\boldsymbol{r}},
\end{equation}
where $\tilde{\boldsymbol{r}}$ denotes the position vector. This choice automatically satisfies the Lorenz gauge condition $\partial_\mu \tilde{A}^\mu = 0$. Henceforth, we suppress the tilde notation for clarity. The interaction between the charged black holes and the external field contributes an additional term to the Lagrangian \cite{Landau:1975pou}
\begin{equation}
    L_{\rm{B}}=\sum_a \frac{1}{c}q_{a}v_{a}^{\mu}A_{\mu}=\sum_a \frac{q_a}{2c}\bm{v_a}\cdot (\bm{B}\times \bm{r_a}),
\end{equation}
where $\boldsymbol{r}_a$ and $\boldsymbol{v}_a$ denote the position and velocity of particle $a$. The complete Lagrangian for the binary system embedded in the external magnetic field becomes:
\begin{equation}\label{eq:total Lagrangian}
    L=L_{\rm{iso}}+L_{\rm{B}}.
\end{equation}
We note that electromagnetic self-interaction terms (proportional to $B^2$) are neglected as they lie beyond the current scope and will be addressed in future work.

The presence of the external field introduces a crucial breaking of translational symmetry. The explicit position dependence in $L_{\text{B}}$ violates spatial translation invariance, resulting in non-conservation of linear momentum. This apparent violation reflects the idealized nature of our uniform field assumption: in a physical scenario, momentum exchange with the field sources would modify the field configuration. By maintaining a stationary uniform field, we implicitly assume that the field sources remain fixed, absorbing the momentum transfer without dynamical response.

The equations of motion follow from the Euler-Lagrange equations
\begin{equation}
    \frac{d}{dt}\frac{\partial L}{\partial\bm{v_{a}}}-\frac{\partial L}{\partial \bm{r_a}}=0,
\end{equation}
where all variables are expressed in three-dimensional form for computational convenience. After extensive algebraic manipulation, the equation of motion for particle 1 yields
\begin{equation}
\begin{aligned}
    & \frac{d}{dt}\frac{\partial L}{\partial\bm{v_{1}}}-\frac{\partial L}{\partial \bm{r_1}}\\
    & =m_{1}\bm{a_1}+\left(\frac{Gm_1 m_2}{r^2}-\frac{q_1 q_2}{r^2}\right)\bm{n}-\frac{q_1}{c}\bm{v_1}\times\bm{B}\\
    & +\frac{1}{c^2}m_{1}(\bm{a_1}\cdot \bm{v_1})\bm{v_1}+\frac{1}{2c^2}m_{1}v_1^2\bm{a_1}\\
    & +\frac{Gm_1 m_2}{2c^2 r}\left[6\bm {a_1}-7\bm{a_2}-(\bm n \cdot \bm{a_2})\bm{n}\right]-\frac{q_1 q_2}{c^2 r^2}(\bm{n}\cdot \bm{v_1})\bm{v_2}\\
    & +\frac{q_1 q_2}{2c^2 r^2}\bm{n}\Big[\frac{4Gq_1 q_2(m_1+m_2)+2G(m_{1}q_{2}^2+m_{2}q_{1}^2)}{r}\\
    & +2v_1 v_2+3(\bm n \cdot \bm{v_2})^{2}-v_2^2\Big]+\frac{q_1 q_2}{2c^2 r}\left[\bm{a_2}+(\bm n \cdot \bm{a_2})\bm{n}\right]\\
    & +\frac{Gm_1 m_2}{2c^2 r^2}\bm{n}\Big[-3(\bm n \cdot \bm{v_2})^{2}+3v_1^2+4v_2^2-8\bm{v_1}\cdot \bm{v_2}\\
    & -\frac{2G(m_1+m_2)}{r}\Big] +\frac{Gm_1 m_2}{c^2 r^2}\bm{v_2}(\bm n \cdot \bm{v_1})\\
    & -\frac{Gm_1 m_2}{2c^2 r^2}(\bm{v_1}-\bm{v_2})\left[6(\bm n \cdot \bm{v_1})-6(\bm n \cdot \bm{v_2})\right]\\
    & =0,
\end{aligned}
\end{equation}
where $r = |\boldsymbol{r}_1 - \boldsymbol{r}_2|$ and $\boldsymbol{n} = (\boldsymbol{r}_1 - \boldsymbol{r}_2)/r$ denotes the unit separation vector.

Since acceleration terms appear at 1PN order, we must perform an order reduction: accelerations in $\mathcal{O}(c^{-2})$ terms are replaced by their Newtonian expressions. This iterative procedure yields the 1PN-accurate acceleration for particle 1

\begin{equation}\label{eq:1PN acceleration}
\begin{aligned}
\bm{a_1} & =-\frac{Gm_2}{r^2}\left(1-\frac{q_1 q_2}{Gm_1 m_2}\right)\bm{n}+\frac{q_1}{m_1}\frac{\bm{v_1}}{c}\times\bm{B}\\
& +\frac{Gm_2}{2c^2 r^2}\bm{n}\Big[\frac{8Gm_2}{r}+\frac{10Gm_1}{r}-\frac{q_1 q_2}{r}(\frac{6}{m_1}+\frac{8}{m_2})\\
& +3(\bm{n}\cdot \bm{v_2})^2+8\bm{v_1}\bm{v_2}-2v_{1}^{2}-4v_{2}^{2}-\frac{q_2}{m_2}r\bm{n}\cdot (\frac{\bm{v_2}}{c}\times \bm{B})\Big]\\
& +\frac{q_1 q_2}{2m_1 c^2 r^2}\bm{n}\Big[v_2^2-v_1^2-2v_1 v_2-3(\bm n \cdot \bm{v_2})^{2}\\ 
&-\frac{4G(m_1+m_2)+2Gm_1-2\frac{q_1 q_2}{m_2}}{r}-\frac{q_2}{m_2}r\bm{n}\cdot (\frac{\bm{v_2}}{c}\times \bm{B})\Big]\\
& +\frac{G(m_{1}q_{2}^2+m_{2}q_{1}^2)}{m_1 c^2}\frac{\bm n}{r^3}-\frac{q_1 q_2}{c^2 m_1 r^2}(\bm{v_1}-\bm{v_2})(\bm{n}\cdot \bm{v_1})\\
& +\frac{Gm_2}{2c^2 r^2}(\bm{v_1}-\bm{v_2})\left[8(\bm n \cdot \bm{v_1})-6(\bm n \cdot \bm{v_2})\right]\\
& -\frac{Gm_2}{2c^2 r}\left(\frac{6q_1}{m_1}\frac{\bm{v_1}}{c}\times \bm{B}-\frac{7q_2}{m_2}\frac{\bm{v_2}}{c}\times \bm{B}\right)\\
& -\frac{q_1 q_2}{2c^2 m_1 r}\frac{q_2}{m_2}\frac{\bm{v_2}}{c}\times \bm{B}-\frac{q_{1}v_{1}^{2}}{2m_1 c^2}\frac{\bm{v_1}}{c}\times\bm{B}+\mathcal{O}(c^{-4}).
\end{aligned}
\end{equation}

The acceleration for particle 2 follows by symmetry under the exchange $1 \leftrightarrow 2$, with the accompanying sign change $\boldsymbol{n} \to -\boldsymbol{n}$. These expressions provide the complete 1PN dynamics necessary for computing orbital evolution in the presence of external magnetic fields.

\section{Solution of the equations of motion to 1PN accuracy}\label{sec:solution}

This section presents a systematic investigation of how external magnetic fields influence the dynamics of binary charged black hole systems and their associated gravitational wave signatures. The magnetic Lorentz force enters at Newtonian order and propagates through all subsequent PN corrections, as higher-order terms inherit contributions from lower-order dynamics, evident in Eq.~(\ref{eq:1PN acceleration}). Combined with the intrinsic charge effects on both dynamics and radiation, we anticipate significant deviations from the evolution of neutral binaries or those without external fields.

The inspiral phase dominates the binary's evolution, comprising the vast majority of its lifetime before merger \cite{Maggiore:2007ulw}. For well-separated black holes, Newtonian dynamics provides sufficient accuracy for modeling trajectories and waveforms \cite{Ioka:2000yb}. However, when the PN parameter $\epsilon$ becomes appreciable yet remains small compared to unity, incorporating 1PN corrections becomes essential. These corrections account for nonlinear gravitational effects and retardation, yielding more accurate trajectories and waveform predictions \cite{Blanchet:2013haa}. We therefore systematically examine both charge and magnetic field contributions, comparing Newtonian and 1PN solutions to quantify the importance of relativistic corrections.

\subsection{Solution at the Newtonian order}

We begin with the Newtonian-order analysis to examine the dominant contributions to the PN corrected accelerations. At this order, the Lagrangian derived from Eq.~(\ref{eq:total Lagrangian}) takes the form
\begin{equation}
\begin{aligned}
    L_{\rm{Newton}}
    & =\frac{1}{2}m_{1}v_{1}^{2}+\frac{1}{2}m_{2}v_{2}^{2}+\frac{Gm_{1}m_{2}}{r}-\frac{q_{1}q_{2}}{r}\\
    & +\frac{q_1}{2c}\bm{v_1}\cdot\left(\bm{B}\times\bm{r_1}\right)+\frac{q_2}{c}\bm{v_2}\cdot\left(\bm{B}\times\bm{r_2}\right),
\end{aligned}
\end{equation}
where constant static energy terms have been omitted as they do not affect the dynamics. This yields the acceleration for particle 1 (with particle 2 obtained via the exchange $1 \leftrightarrow 2$)
\begin{equation}\label{eq:Newtonian acceleration}
    \bm{a_1}=-\frac{Gm_2}{r^2}\left(1-\frac{q_1 q_2}{Gm_1 m_2}\right)\bm{n}+\frac{q_1}{m_1}\frac{\bm{v_1}}{c}\times\bm{B}.
\end{equation}

Without the magnetic field, total linear momentum would be conserved at this order. However, examining the accelerations in Eq.~(\ref{eq:Newtonian acceleration}), we find
\begin{equation}
    m_1 \bm{a_1}+m_2 \bm{a_2}=q_1\frac{\bm{v_1}}{c}\times\bm{B}+q_2\frac{\bm{v_2}}{c}\times\bm{B}.
\end{equation}
This represents the time derivative of the total linear momentum at Newtonian order
\begin{equation}
    m_1 \bm{a_1}+m_2 \bm{a_2}=\frac{d}{dt}(m_1 \bm{v_1}+m_2 \bm{v_2})=\frac{d}{dt}(\bm{p_1}+\bm{p_2})=\frac{d}{dt}\bm P,
\end{equation}
where $\bm{p_a}=\partial L_{\rm{Newton}}/\partial \bm{v_a}$ denotes the linear momentum of particle $a$. Then we introduce the vector
\begin{equation}
    \bm Z =\frac{m_1 \bm{r_1}+m_2 \bm{r_2}}{m_1+m_2},
\end{equation}
as the position of the center-of-mass (CoM). In the absence of $\bm B$, in the CoM frame we always have $d\bm P/dt=0$. Thus it is convenient to set $\bm Z=0$ in the CoM frame to investigate the dynamics of an isolated binary system. As long as $\bm B\neq 0$, the total linear momentum is generally not conserved, i.e. the second time derivative of $\bm Z$ is generally not 0. Consequently, there is a secular drift of the CoM position. Taking the cross product of $m_1 \bm{a_1}$ with $\bm{r_1}$ and $m_2 \bm{a_2}$ with $\bm{r_2}$ yields
\begin{equation}
\begin{aligned}
    & \bm{r_1}\times m_1 \bm{a_1}+\bm{r_2}\times m_2 \bm{a_2}\\
    & =\frac{d}{dt}(\bm{r_1}\times m_1 \bm{v_1}+\bm{r_2}\times m_2 \bm{v_2})=\frac{d}{dt}\bm J,
\end{aligned}
\end{equation}
where $\bm J$ represents the total angular momentum, which reduces to the orbital angular momentum in the CoM frame. Using Eq.~(\ref{eq:Newtonian acceleration}), we obtain
\begin{equation}\label{eq:time derivative of J}
\begin{aligned}
    \frac{d}{dt}\bm J &= q_1\bm{r_1}\times(\frac{\bm{v_1}}{c}\times\bm{B})+q_2\bm{r_2}\times(\frac{\bm{v_2}}{c}\times\bm{B})\\
    &= q_1 \left[(\bm{r_1}\cdot\bm{B})\frac{\bm{v_1}}{c}-(\bm{r_1}\cdot\frac{\bm{v_1}}{c})\bm{B}\right]\\
    &+q_2 \left[(\bm{r_2}\cdot\bm{B})\frac{\bm{v_2}}{c}-(\bm{r_2}\cdot\frac{\bm{v_2}}{c})\bm{B}\right].
\end{aligned}
\end{equation}

Angular momentum is therefore not conserved in general, a direct consequence of the uniform stationary magnetic field. Conservation occurs only in the special case where the orbital plane is perpendicular to $\bm B$ and the orbit is circular. Consequently, initially planar orbits evolve secularly into three-dimensional trajectories.

To analyze this evolution quantitatively, we consider the time-averaged derivative of Eq.~(\ref{eq:time derivative of J}). For any bounded function $f(t)$, the time-averaged derivative over period $T$ is given by \cite{Landau:1975pou}
\begin{equation}
    \overline{\frac{df}{dt}}=\frac{1}{T}\int^{t}_{0} \frac{df}{dt'}dt'=\frac{f(t)-f(0)}{T}.
\end{equation}
In the limit $T\rightarrow +\infty$, we have $\overline{\frac{df}{dt}} \rightarrow 0$ for bounded motion. Thus, terms involving $\bm{r}\cdot \bm{v}$ average to zero, yielding
\begin{equation}
\begin{aligned}
    \overline{\frac{d}{dt}\bm J} 
    &= \sum q_a \overline{(\bm{r_a}\cdot\bm{B})\frac{\bm{v_a}}{c}}\\
    &= \frac{1}{2c}\sum q_a\left[\overline{\bm{v_a}(\bm{r_a}\cdot\bm{B})}-\overline{\bm{r_a}(\bm{v_a}\cdot\bm{B})}\right]\\
    & +\frac{1}{2c}\overline{\frac{d}{dt}\left[\bm{r_a}(\bm{r_a}\cdot\bm{B})\right]}=\overline{\bm{d_m}}\times \bm{B},
\end{aligned}
\end{equation}
where $\bm{d_m}=\frac{1}{2c}\sum q_a \bm{r_a}\times \bm{v_a}$ defines the magnetic moment, with total time derivatives vanishing upon averaging. Expressing the magnetic moment as $\bm{d_m}=\frac{1}{2c}\sum (q_a/m_a) \bm{r_a}\times m_a\bm{v_a}=(q_1/2cm_1)\bm{J}+\frac{1}{2c}(q_{2}/m_{2}-q_{1}/m_{1})\bm{r_2}\times m_2\bm{v_2}$ gives
\begin{equation}\label{eq:time average of torque}
\begin{aligned}
    \overline{\frac{d}{dt}\bm J}
    & =\overline{\bm{d_m}}\times \bm{B}=\frac{q_1}{2cm_1}\overline{\bm{J}}\times \bm{B}\\
    & +\frac{m_2}{2c}(\frac{q_{2}}{m_{2}}-\frac{q_{1}}{m_{1}})\overline{\left[(\bm{r_2}\cdot \bm{B})\bm{v_2}-(\bm{v_2}\cdot \bm{B})\bm{r_2}\right]}.
\end{aligned}
\end{equation}

\begin{figure}[t]
    \centering
    \includegraphics[width=0.4\linewidth]{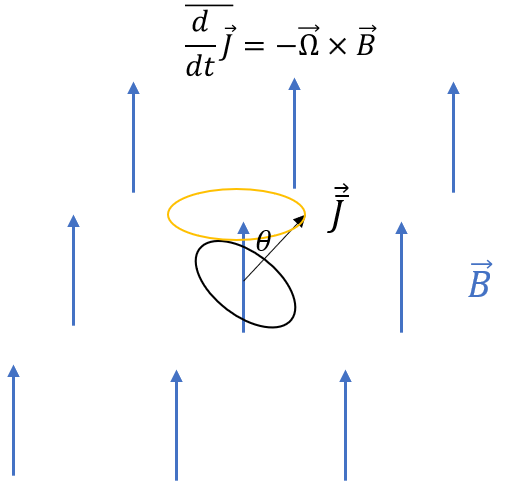}
    \hspace{0.1\linewidth}
    \includegraphics[width=0.4\linewidth]{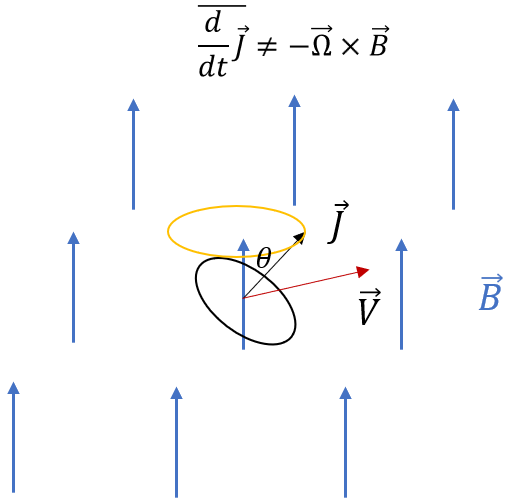}
    \caption{The schematic diagram of the evolution of the orbits. The left panel represent the case with $q_1/m_1=q_2/m_2$ and the right is for $q_1/m_1\neq q_2/m_2$. $\vec V$ is the velocity of CoM which is not conserved.}
    \label{fig:schematic of orbital motion}
\end{figure}

\begin{figure}
    \centering
    \includegraphics[width=0.45\linewidth]{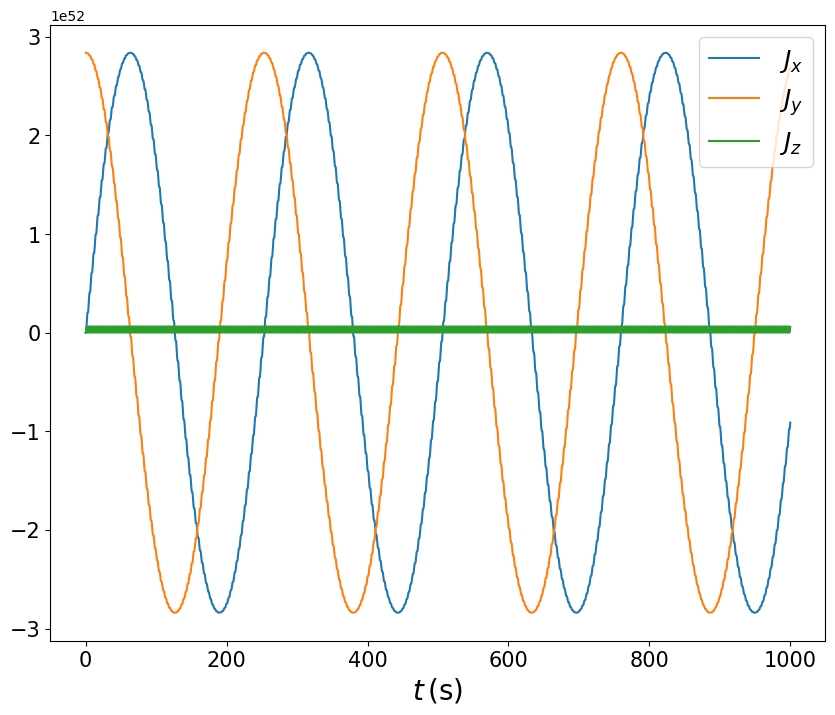}
    \hspace{0.01\linewidth}
    \includegraphics[width=0.45\linewidth]{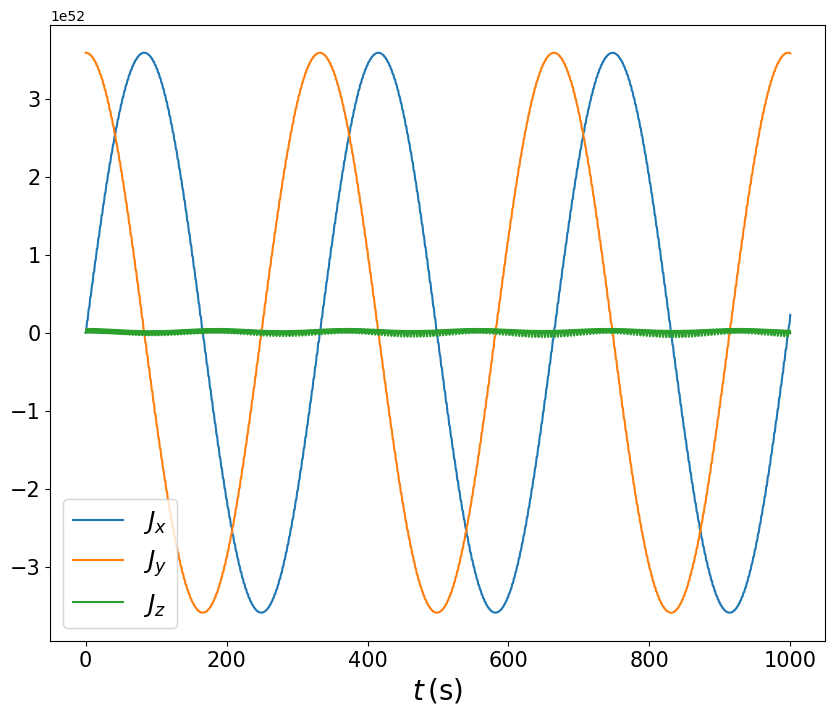}
    \caption{Cartesian components of $\bm J$ for different charge-to-mass ratios at Newtonian order. Physical parameters: $m_1=15M_{\odot}$, $q_1=\sqrt{G}\nu_{1}m_1$ with $\nu_1=0.72$; $m_2=20M_{\odot}$, $q_2=\sqrt{G}\nu_{2}m_2$ with $\nu_2=0.72$ (left panel) and $\nu_2=0.32$ (right panel).}
    \label{fig:angular momentum at 0PN}
\end{figure}

When $q_2/m_2=q_1/m_1$, the magnitude of $\bm J$ remains constant while its direction precesses with angular velocity $\bm \Omega=(q_1/2cm_1)\bm{B}$ \cite{Landau:1975pou}. Under this condition, if the initial total momentum $\bm P$ vanishes, it remains zero throughout the evolution. The evolution of the total linear and angular momentum is shown in Figure~\ref{fig:schematic of orbital motion} and Figure~\ref{fig:angular momentum at 0PN} illustrates the angular momentum evolution for different charge-to-mass ratios.

For numerical integration, we orient the magnetic field along the $Z$ axis. Initial velocities lie in the $X$-$Z$ plane to ensure three-dimensional orbital evolution. Following Keplerian dynamics, the initial positions are
\begin{equation}
     \bm{r_1}=\left(a(1-e)\frac{m_2}{M},0,0\right), \quad
     \bm{r_2}=\left(-a(1-e)\frac{m_1}{M},0,0\right),
\end{equation}
where $M=m_1+m_2$ is the total mass, $a$ the semi-major axis, and $e$ the eccentricity. The initial velocities are
\begin{equation}
    \bm{v_1}=-v_0\frac{m_2}{M}(0,0,1), \quad
    \bm{v_2}=v_0\frac{m_1}{M}(0,0,1),
\end{equation}
with $v_0=\sqrt{GM(1-\frac{q_1 q_2}{Gm_1 m_2})\frac{(1+e)}{a(1-e)}}$ chosen to ensure zero initial linear momentum. For our analysis, we set $e=0$ and $a=240GM/c^{2}$.

Denoting {$\nu_1 \equiv q_1/\sqrt{G}m_1$ and $\nu_2\equiv q_2/\sqrt{G}m_2$} as the charge-to-mass ratios, Figure~\ref{fig:angular momentum at 0PN} displays results for a strong magnetic field $\bm B=8.0\times 10^{12}\,\mathrm{G}$ \cite{magnetic_field_of_neutron_star} to clearly demonstrate the evolution. The rapid oscillations arise from the orbital motion and average to zero. When $\nu_1 \neq \nu_2$, the $Z$-component magnitude exhibits significant secular variation. 

Since the Newtonian Lagrangian contains no explicit time dependence, the total energy $E$ is conserved, as the magnetic Lorentz force performs no work. However, the secular evolution of both the CoM position and orbital angular momentum direction drives the transition to three-dimensional orbits, precluding analytical solutions. We therefore proceed with numerical integration of the equations of motion at both Newtonian and higher PN orders in the following subsection.

\subsection{Solution at the first post-Newtonian order}

We now examine the 1PN-order corrections to the equations of motion and their implications for gravitational waveforms.

Beginning with the case without an external magnetic field, we observe that conservation laws depend on the order of approximation considered. The total linear momentum takes the form
\begin{equation}
\begin{aligned}
    \bm{P} &\equiv\bm{p_1}+\bm{p_2}= \frac{\partial L_{\rm{iso}}}{\partial \bm{v_1}}+\frac{\partial L_{\rm{iso}}}{\partial \bm{v_2}}= m_1 \bm{v_1}+m_2 \bm{v_2}\\
    & +\frac{1}{c^2}\Big[\frac{1}{2}m_1 v_{1}^{2}\bm{v_1}+\frac{1}{2}m_2 v_{2}^{2}\bm{v_2}\\
    & +\frac{Gm_1 m_2}{2r}\left(-\bm{v_1}-\bm{v_2}- (\bm{n}\cdot\bm{v_1})\bm{n}-(\bm{n}\cdot\bm{v_2})\bm{n}\right)\\
    & +\frac{q_1 q_2}{2r}(\bm{v_1}+\bm{v_2}+(\bm{n}\cdot\bm{v_1})\bm{n}+(\bm{n}\cdot\bm{v_2})\bm{n})\Big].
\end{aligned}
\end{equation}

At 1PN accuracy, conservation of total linear momentum reveals a deviation in the CoM position from its Newtonian definition, analogous to the uncharged case \cite{Blanchet:1998vx}. Consequently, the relation $m_1\bm{r_1}+m_2\bm{r_2}=0$ is not applicable to determine the CoM frame at this order, even though there is no background magnetic field. Therefore, we avoid specifying a particular coordinate system in our numerical integration.

In the presence of an external magnetic field, energy remains the sole conserved quantity, as at Newtonian order. The 1PN-accurate accelerations become too complex for analytical treatment, necessitating numerical analysis.

Energy conservation holds only to 1PN order, specifically
\begin{equation}
    \frac{dE}{dt}=\mathcal{O}(c^{-4}),
\end{equation}
due to the order-reduction procedure applied to the accelerations. This variation is consistent with $\epsilon^{2}$ scaling for the orbital parameters considered.

\begin{figure}[t]
    \centering
    \includegraphics[width=0.650\linewidth]{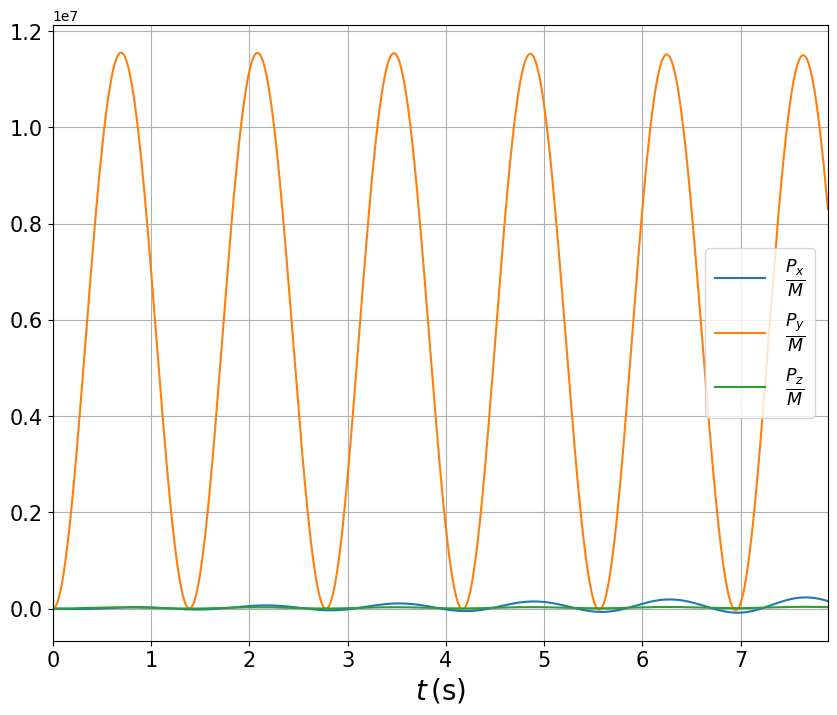}
    \caption{Three components of the Newtonian-order total linear momentum evolution. Physical parameters: $m_1=15M_{\odot}$, $q_1=\sqrt{G}\nu_{1}m_1$ with $\nu_1=0.52$; $m_2=20M_{\odot}$, $q_2=\sqrt{G}\nu_{2}m_2$ with $\nu_2=-0.48$. The magnetic field strength is $\bm B=5.0\times 10^{12}\,\mathrm{G}$. The linear momenta are rescaled by the total mass $M$.}
    \label{fig:1PN momentum}
\end{figure}

Figure~\ref{fig:1PN momentum} illustrates the evolution of the Newtonian-order total linear momentum since the 1PN correction is small and we mainly focus on the effect of magnetic field. The momentum exhibits secular growth, particularly along the third coordinate direction (the $Y$-axis in our configuration). This secular behavior arises from CoM motion: at Newtonian order, the external magnetic field prevents separation of overall and relative motions. The resulting CoM acceleration induces a magnetic Lorentz force on the total charge $Q=q_1+q_2$, effectively mimicking a point mass $M$ with charge $Q$ moving in the external field. This mechanism drives the observed increase in the $X$-component of total linear momentum $P_x$, as shown in Figure~\ref{fig:1PN momentum}.

\subsection{Comparison of the results}

To assess the influence of the magnetic field, we examine two distinct scenarios: orbital motions with and without an external magnetic field, evaluated at both Newtonian and 1PN orders.

\begin{figure}[t]
    \centering
    \includegraphics[width=0.45\linewidth]{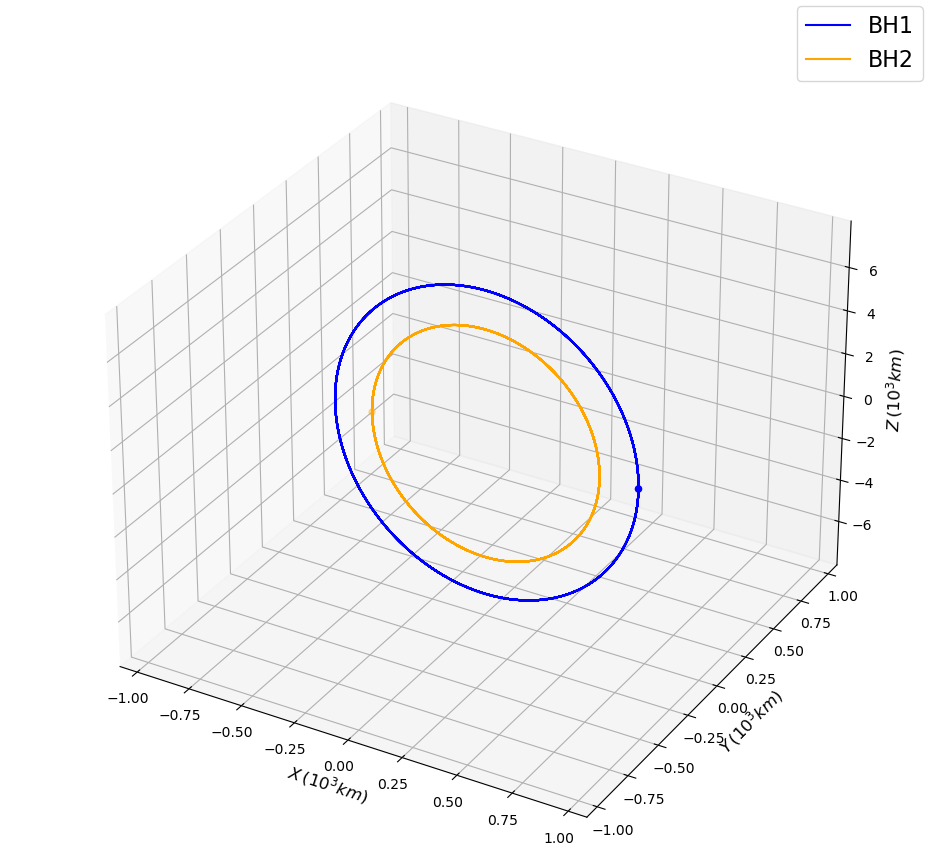}
    \hspace{0.01\linewidth}
    \includegraphics[width=0.45\linewidth]{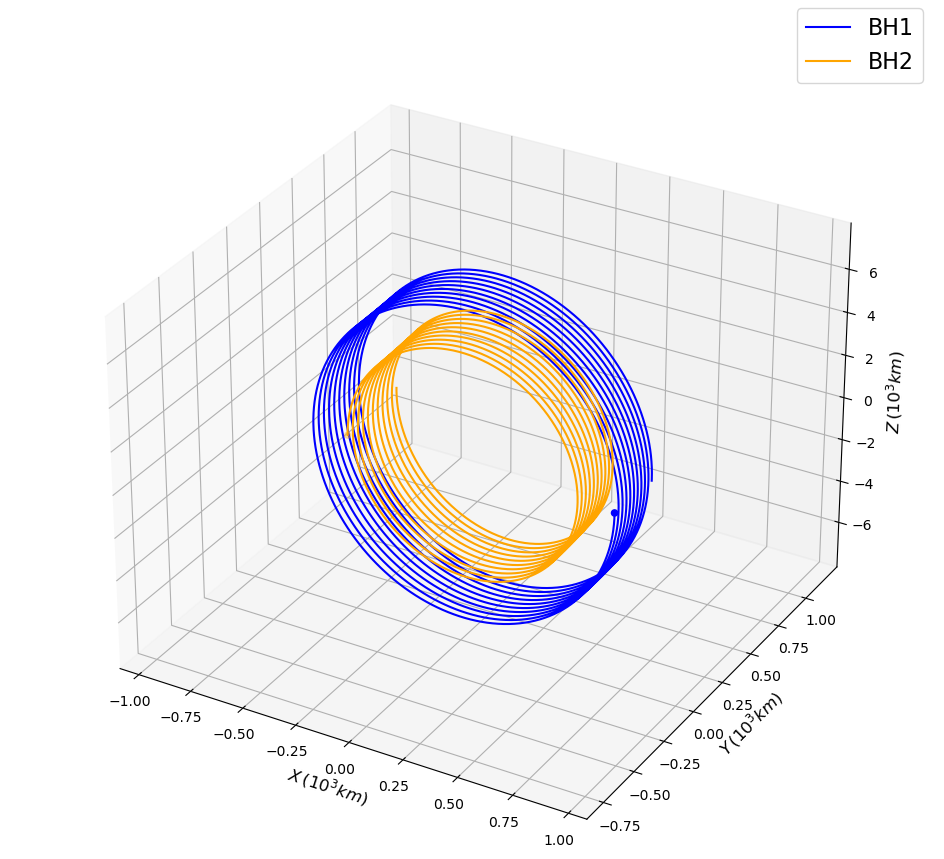}
    \caption{Comparison of trajectories with (right) and without (left) magnetic field at Newtonian order. Physical parameters: $m_1=15M_{\odot}$, $q_1=\sqrt{G}\nu_{1}m_1$ with $\nu_1=0.52$; $m_2=20M_{\odot}$, $q_2=\sqrt{G}\nu_{2}m_2$ with $\nu_2=-0.48$.}
    \label{fig:0PN trajectory comparison}
\end{figure}

Figure~\ref{fig:0PN trajectory comparison} presents the numerically integrated trajectories for both cases at Newtonian order. Without an external magnetic field, the orbit remains circular and confined to the $X$-$Z$ plane, consistent with the uncharged case. In contrast, the presence of a magnetic field induces pronounced secular motion of the CoM, which manifests as part of a magnetic-Lorentz trajectory—effectively corresponding to a mass $M$ carrying total charge $Q$ with initial velocity along the $Y$-axis in a uniform magnetic field.

\begin{figure}[t]
    \centering
    \includegraphics[width=0.45\linewidth]{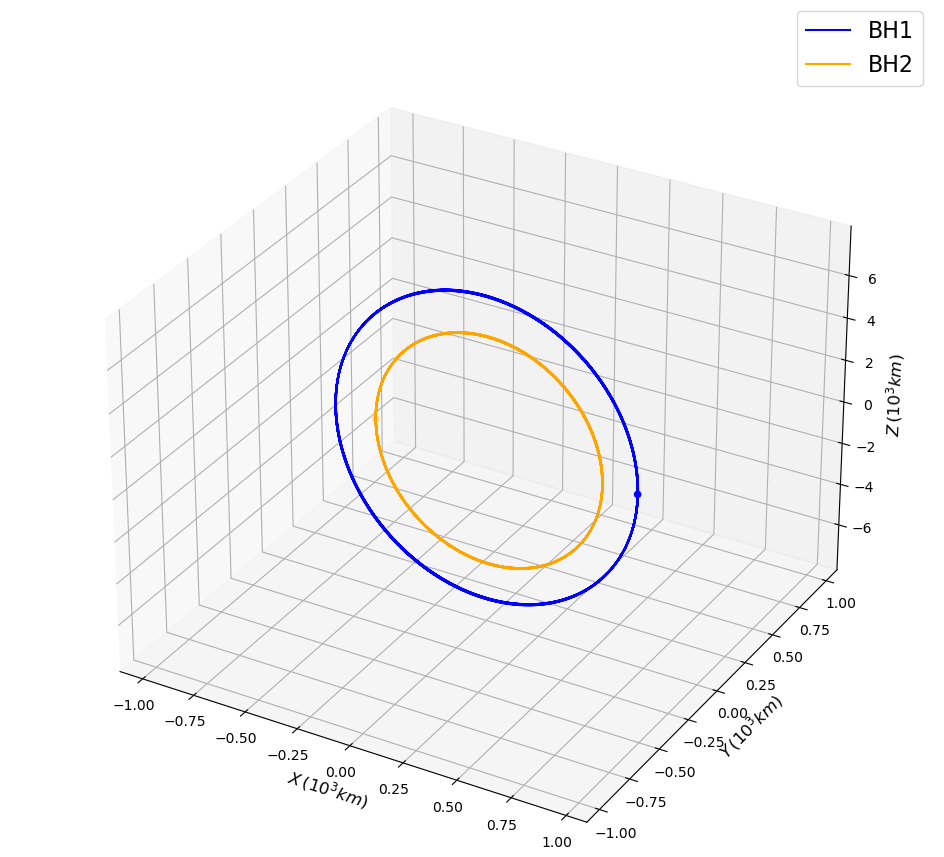}
    \hspace{0.01\linewidth}
    \includegraphics[width=0.45\linewidth]{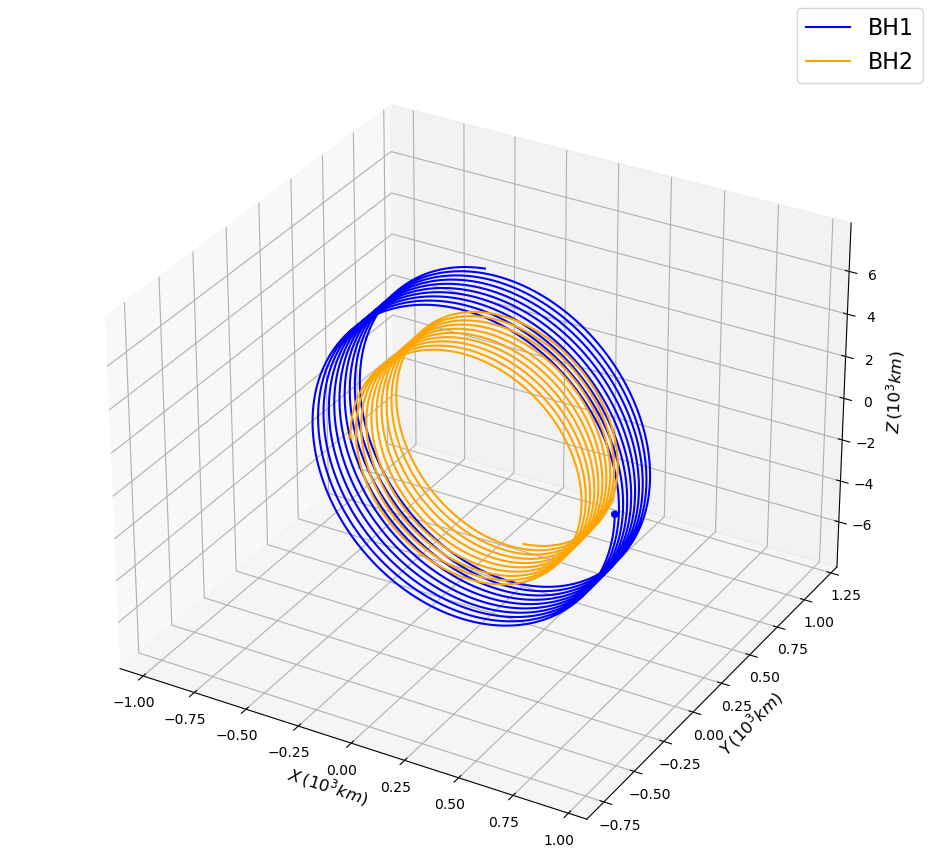}
    \caption{Comparison of trajectories with (right) and without (left) magnetic field at 1PN order. Physical parameters identical to Figure~\ref{fig:0PN trajectory comparison}.}
    \label{fig:1PN trajectory comparison}
\end{figure}

At 1PN order, the corresponding trajectories appear in Figure~\ref{fig:1PN trajectory comparison}. Given the perturbative nature of the expansion parameter $\epsilon$, the 1PN corrections remain small, preventing significant deviation from Newtonian trajectories. Nevertheless, as demonstrated below, PN effects remain discernible at this order.

\begin{figure}[t]
    \centering
    \includegraphics[width=0.45\linewidth]{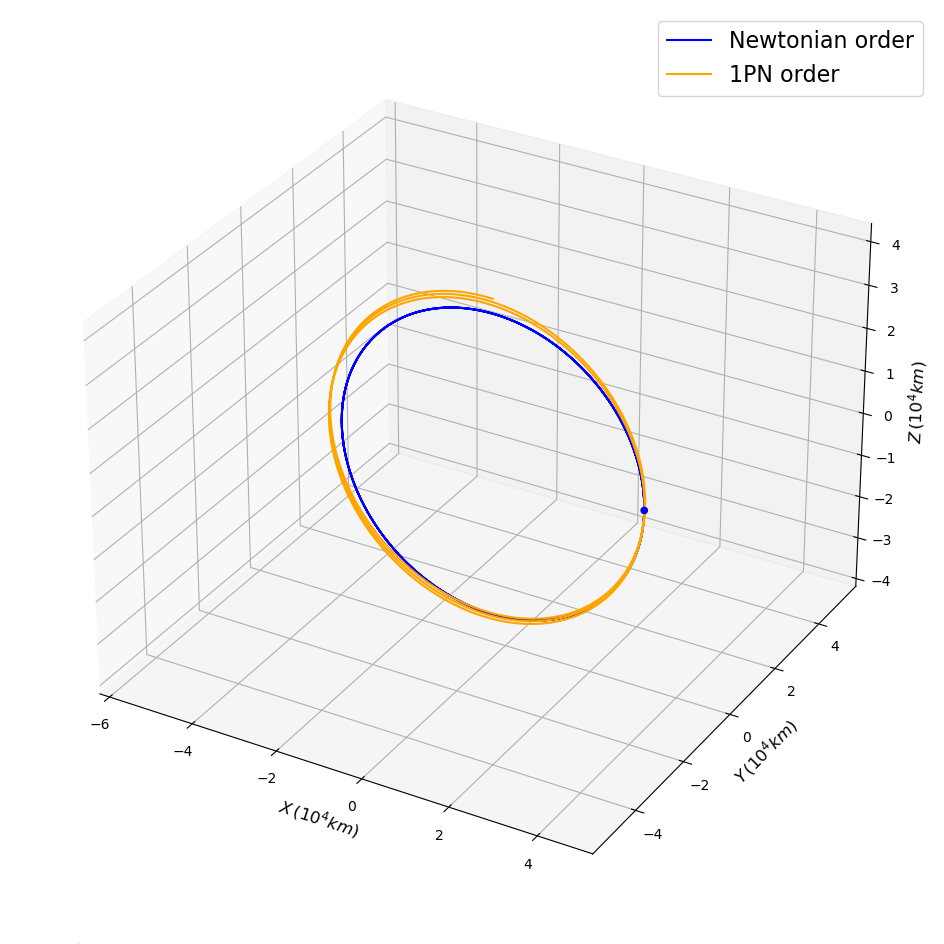}
    \hspace{0.01\linewidth}
    \includegraphics[width=0.45\linewidth]{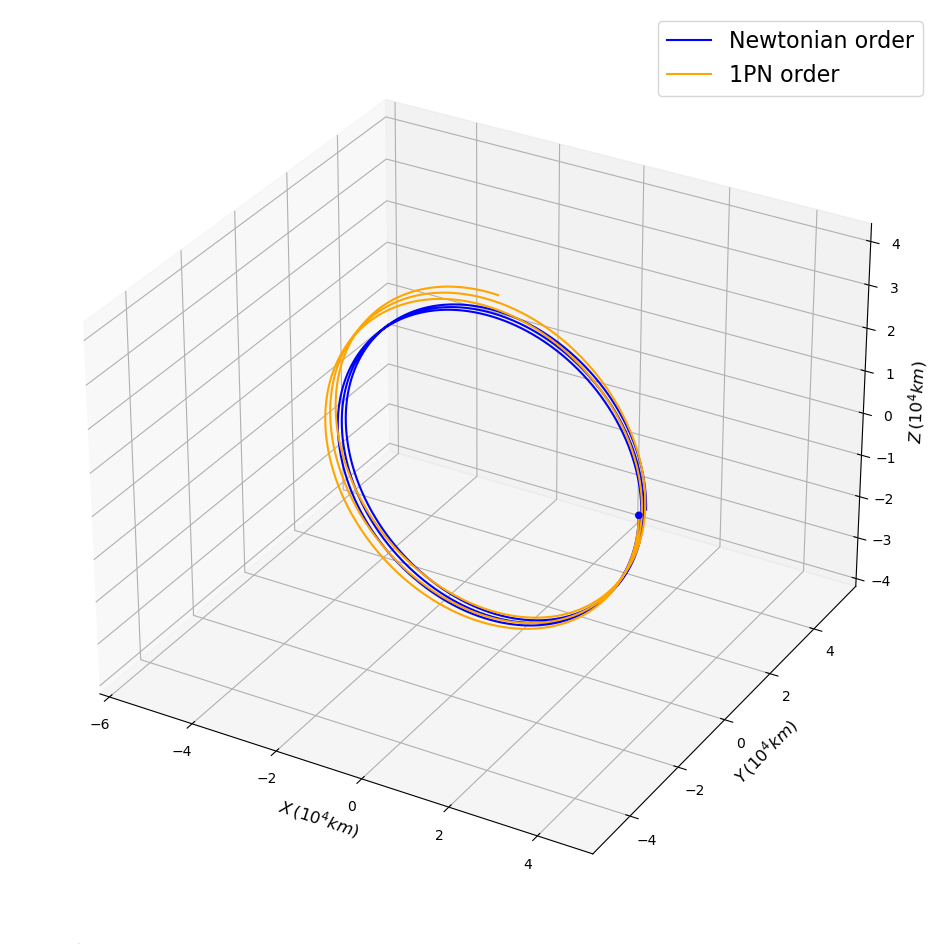}
    \caption{Comparison of trajectories at Newtonian and 1PN orders, for the case with (right) and without (left) magnetic field. Physical parameters identical to Figure~\ref{fig:0PN trajectory comparison}. The time span corresponds to two Keplerian orbital periods for clarity.}
    \label{fig:compare 0PN and 1PN}
\end{figure}

Figure~\ref{fig:compare 0PN and 1PN} illustrates the 1PN acceleration corrections for the first black hole. Our analysis reveals that elliptic orbits exhibit larger 1PN corrections compared to circular orbits, attributable to the additional contributions from velocity-dependent dot products in Eq.~(\ref{eq:1PN acceleration}). Consequently, we focus on elliptic orbits with eccentricity $e=0.1$ to better characterize the interplay between 1PN corrections and magnetic field effects.

The comparison spans only two Newtonian orbital periods to clearly visualize the 1PN corrections. The orbital deviations prove substantial, consistent with the expected PN scaling. Notably, despite treating the magnetic field as external in the Lagrangian formulation, it contributes significantly to the 1PN corrections through its influence on the lower-order accelerations that enter the higher-order terms.

Our comparative analysis yields several key insights regarding the dynamics of charged binary systems:
\begin{enumerate}
    \item In the absence of magnetic fields, orbital motion remains confined to a plane at all PN orders, preserving the fundamental geometric structure of gravitational two-body dynamics.
    \item External magnetic fields fundamentally alter the orbital geometry by inducing three-dimensional trajectories, even when the initial configuration is planar. This breaking of planar symmetry represents a qualitative departure from pure gravitational dynamics.
    \item The CoM exhibits secular drift that becomes increasingly pronounced over multiple orbital periods, reflecting the cumulative effect of the magnetic Lorentz force acting on the net charge of the system.
    \item While 1PN corrections quantitatively modify the orbital shape and introduce additional precession effects, they preserve the qualitative features induced by the magnetic field, indicating that magnetic effects dominate the large-scale orbital structure.
\end{enumerate}

These findings demonstrate that external magnetic fields introduce fundamentally new dynamical features that cannot be captured by gravitational effects alone, emphasizing the importance of electromagnetic contributions in astrophysical environments where charged compact objects may exist.

\section{Influence of magnetic field on gravitational waveforms}\label{sec:waveform}
The orbital dynamics analyzed in the previous section have direct implications for the gravitational radiation emitted by charged binary systems. The presence of an external magnetic field not only modifies the orbital trajectories but also fundamentally alters the characteristics of the emitted gravitational waves. This modification occurs through two primary mechanisms: the magnetic field-induced precession of the orbital plane, which affects the waveform polarization content, and the secular evolution of the system's quadrupole moment, which influences the amplitude modulation.

In this section, we investigate how these magnetic field effects manifest in the gravitational waveforms. We begin by computing the waveforms in the time domain using the quadrupole formalism, then examine their frequency-domain characteristics to identify distinctive signatures that could potentially be observed by current and future gravitational wave detectors.

\subsection{Gravitational waveforms in time domain}

Having obtained the binary trajectories from the 1PN equations of motion, we now compute the corresponding gravitational waveforms. Following the quadrupole formalism \cite{Maggiore:2007ulw}, the plus and cross polarization components are given by
\begin{equation}
\begin{aligned}\label{eq:waveform quadrupole formula}
    h_{+}(t;\theta,\phi)
    & =\frac{1}{L}\frac{G}{c^4}\Big[\Ddot{M}_{11}(\cos^2\phi-\sin^2\phi \cos^2\theta)\\
    & +\Ddot{M}_{22}(\sin^2\phi-\cos^2\phi \cos^2\theta)-\Ddot{M}_{33}\sin^2\theta\\
    & +\Ddot{M}_{12}\sin2\phi(1+\cos^2\theta)+\Ddot{M}_{13}\sin\phi \sin2\theta\\
    & +\Ddot{M}_{23}\cos\phi \sin2\theta\Big],
\end{aligned}
\end{equation}

\begin{equation}
\begin{aligned}
    h_{\times}(t;\theta,\phi)
    & =\frac{1}{L}\frac{G}{c^4}\Big[(\Ddot{M}_{11}-\Ddot{M}_{22})\sin2\phi \cos\theta\\
    & +2\Ddot{M}_{12}\cos2\phi\cos\theta-2\Ddot{M}_{13}\cos\phi \sin\theta\\
    & +2\Ddot{M}_{23}\sin\phi \sin\theta\Big],
\end{aligned}
\end{equation}
where $M_{ij}=\sum_a m_a x_i x_j$ denotes the mass quadrupole moment tensor of the binary system, the overdot represents time derivatives, $L$ is the luminosity distance to the observer, and the angles $(\theta,\phi)$ parametrize the propagation direction $\bm{n}=(\sin\theta \sin\phi,\sin\theta \cos\phi,\cos\theta)$. Through the trajectory-dependent quadrupole moment, the magnetic field effects become encoded in the gravitational waveforms.

\begin{figure}[t]
    \centering
    \includegraphics[width=0.45\linewidth]{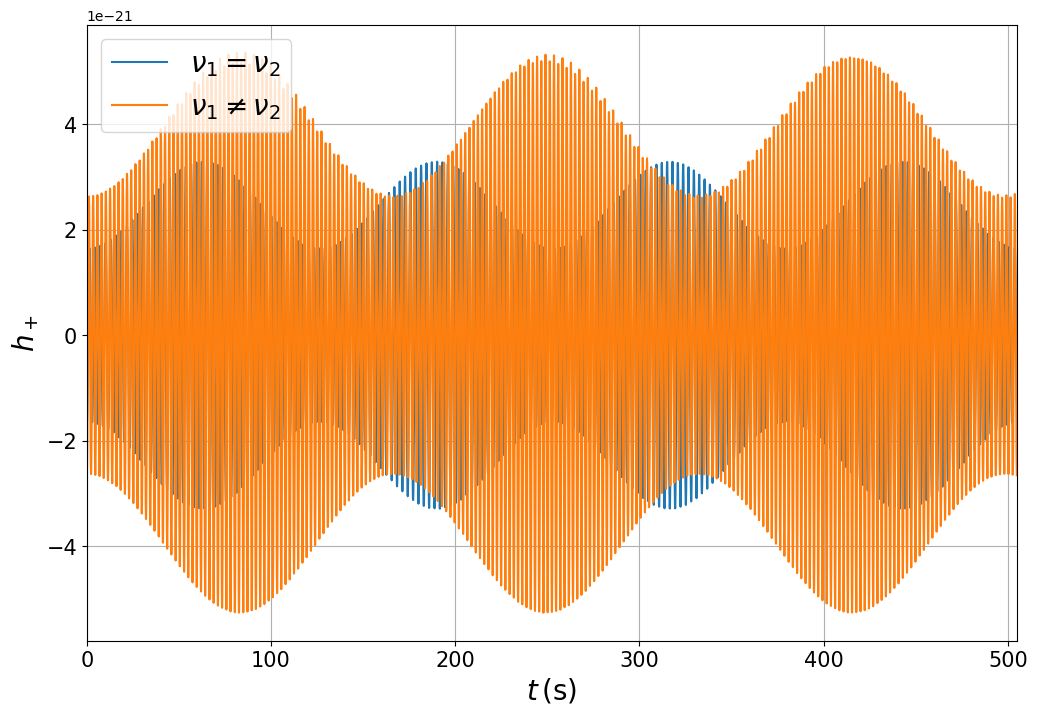}
    \hspace{0.01\linewidth}
    \includegraphics[width=0.45\linewidth]{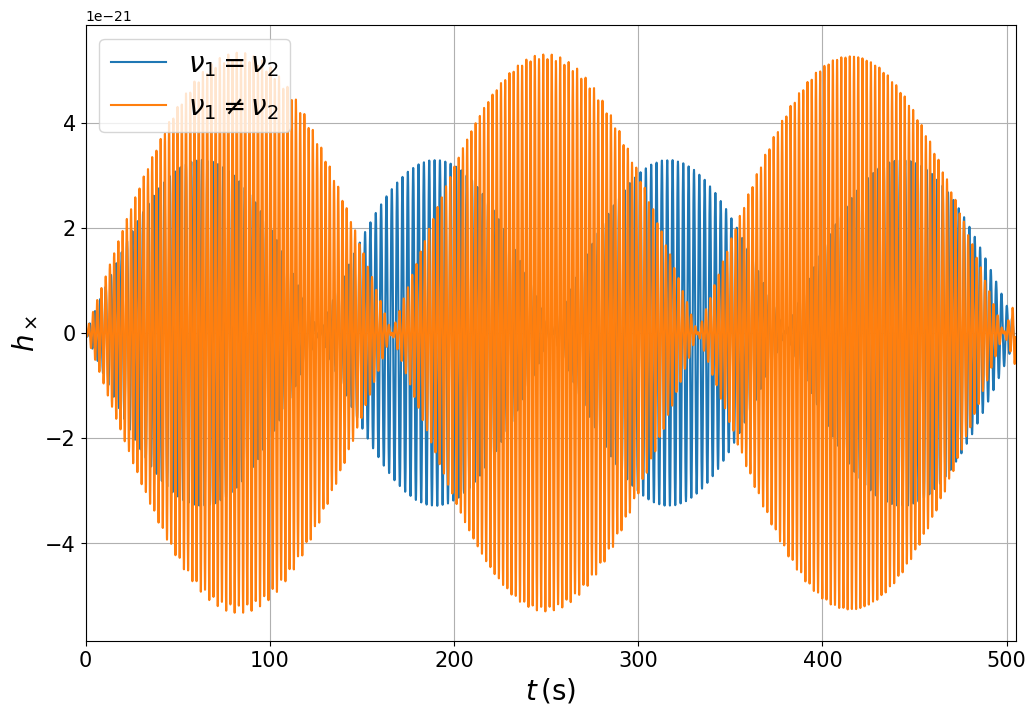}
    \caption{Gravitational waveform polarizations in the time domain. Left panel: plus polarization component. Right panel: cross polarization component. Different colors distinguish systems with identical ($\nu_1=\nu_2$) and different ($\nu_1\neq\nu_2$) charge-to-mass ratios. Physical parameters and initial conditions match those in Figure~\ref{fig:angular momentum at 0PN}. The observation direction is $\bm{n}=(1,0,0)$ along the $X$-axis, with luminosity distance $L=1\,\rm{Mpc}$.}
    \label{fig:waveforms comparison}
\end{figure}

Figure~\ref{fig:waveforms comparison} presents the plus and cross polarization components for systems with and without magnetic fields, computed without accounting for 1PN correction and energy dissipation. The magnetic field induces a striking enhancement in the cross polarization amplitude, exhibiting periodic behavior when $\nu_1=\nu_2$ and quasi-periodic modulation when $\nu_1\neq \nu_2$. This phenomenon arises from the precession of the orbital angular momentum around the magnetic field direction, as described by Eq.~(\ref{eq:time average of torque}). 

For our chosen observation direction along the $X$-axis, the system alternates between face-on and edge-on configurations relative to the observer. The edge-on orientation yields minimal cross polarization, while the face-on configuration maximizes the cross component amplitude. This periodic alternation between geometric configurations produces the characteristic amplitude modulation visible in the waveforms, representing a unique signature of magnetically influenced binary dynamics.

\subsection{Radiative dissipation}

The emission of gravitational and electromagnetic radiation drives the orbital decay of the binary system, causing the two black holes to inspiral toward each other. This dynamical evolution is governed by radiation reaction forces, which appear at odd powers of $v/c$ in the PN expansion due to their intrinsic breaking of time-reversal symmetry. We incorporate the lowest-order radiation reaction terms for both electromagnetic and gravitational contributions.

For electromagnetic radiation, charge conservation dictates that the leading-order radiation reaction appears at $\mathcal{O}(c^{-3})$, corresponding to electric dipole radiation. Following Poisson and Will \cite{poisson2014gravity}, the acceleration for particle $a$ due to electric dipole radiation takes the form
\begin{equation}
    \bm{a_{\rm{1.5PN}}}=\frac{2}{3c^3}\frac{q_{a}}{m_a}\bm{I_e}^{(3)},
\end{equation}
where $\bm{I}_e=\sum_a q_a \bm{x_a}$ denotes the electric dipole moment of the system, with superscript $(n)$ indicating the $n$-th time derivative.

In contrast, gravitational radiation reaction first appears at 2.5PN order, as conservation of total mass and linear momentum eliminates all $\mathcal{O}(c^{-3})$ contributions. We derive the corresponding acceleration from the Burke-Thorne potential \cite{Maggiore:2007ulw}
\begin{equation}
    V_{\rm{reac}}=-\frac{G}{5c^5}x_i x_j M_{ij}^{(5)},
\end{equation}
yielding the acceleration:
\begin{equation}
    \bm{a_{2.5\rm{PN}}}=-\frac{2G}{5c^5}x_j M_{ij}^{(5)}.
\end{equation}

Combining these radiation reaction terms with the conservative dynamics, we numerically integrate the equations of motion for inspiraling charged black holes and subsequently compute the gravitational waveforms using Eq.~(\ref{eq:waveform quadrupole formula}).

To systematically investigate the effects of charges and magnetic fields, we examine three representative cases: (1) a system with both charges and magnetic field, (2) a charged system without magnetic field, and (3) an uncharged system without magnetic field, hereafter denoted as Cases 1, 2, and 3, respectively.

\begin{figure}
    \centering
    \includegraphics[width=0.45\linewidth]{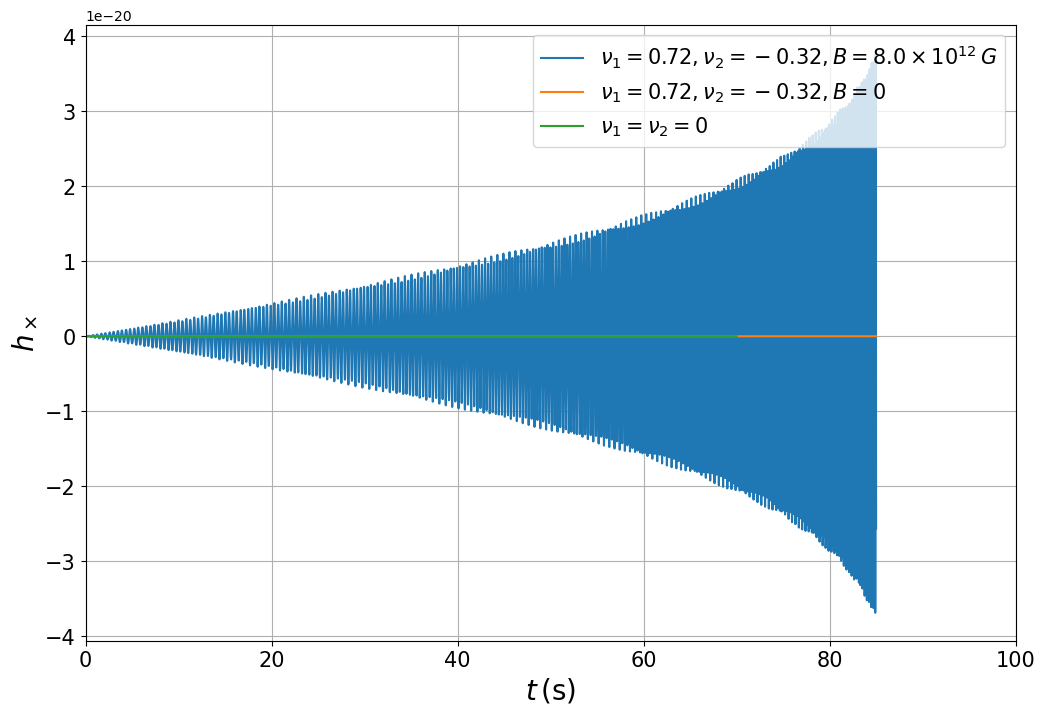}
    \includegraphics[width=0.45\linewidth]{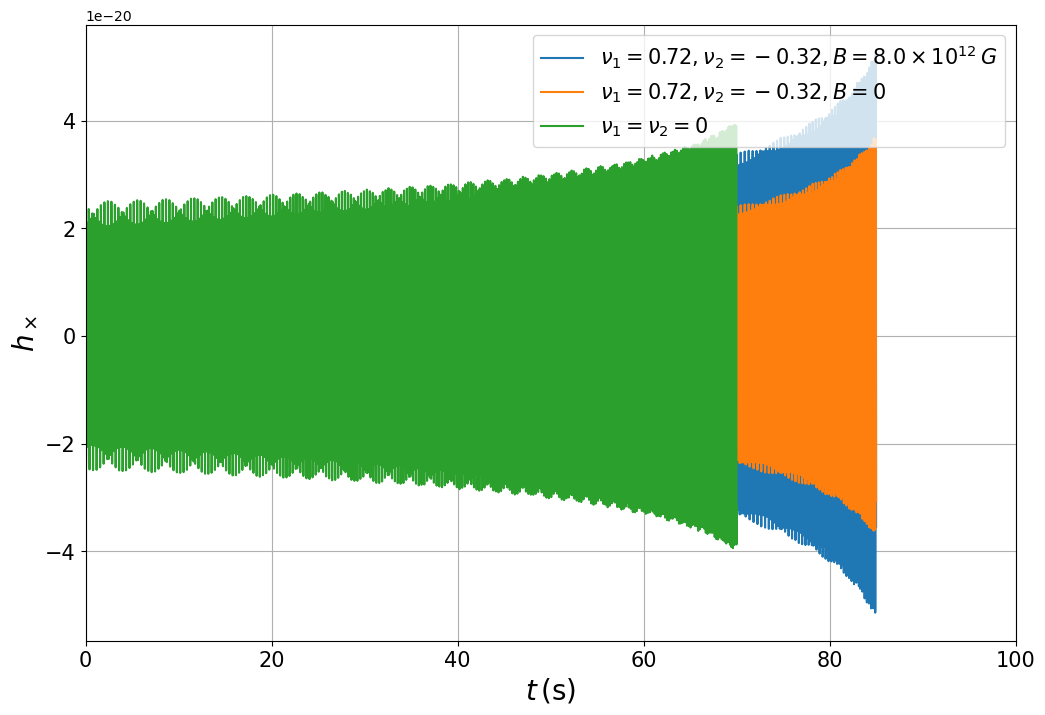}
    \includegraphics[width=0.45\linewidth]{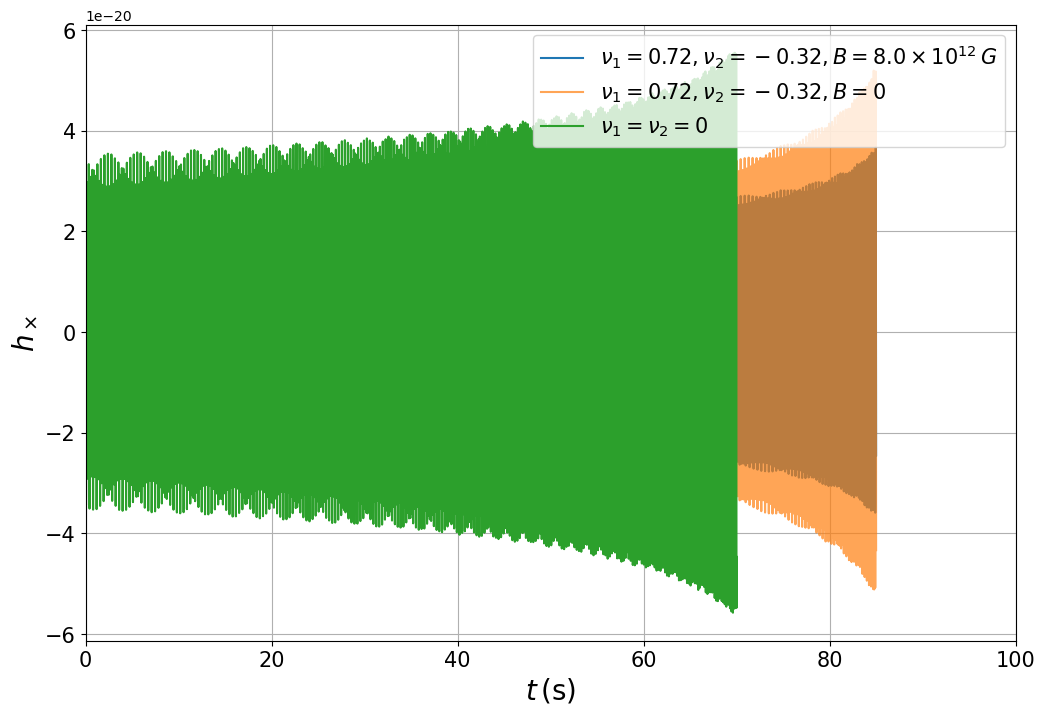}
    \includegraphics[width=0.45\linewidth]{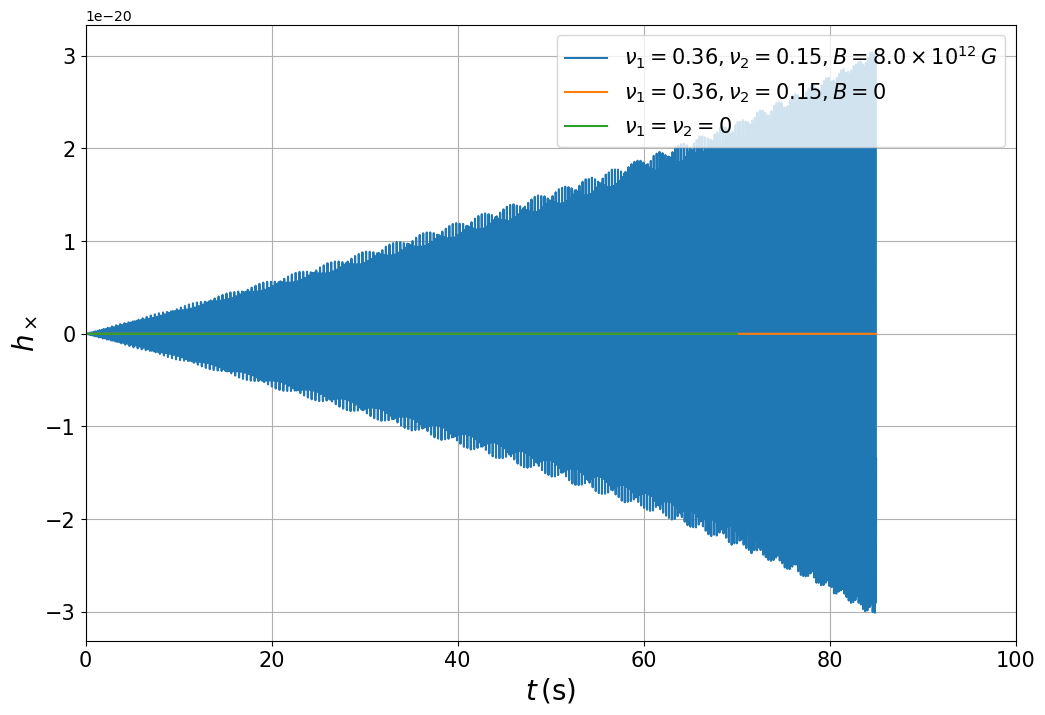}
    \caption{Gravitational waveform cross polarizations including 1PN corrections and radiation reaction, with different observation orientation. The top left panel: along the $X$-axis. Top right panel: $\pi/4$ to the $X$-axis. Bottom left: $Y$-axis. Bottom right: $X$-axis but for different $\nu_1,\nu_2$. Physical parameters: $m_1=15M_{\odot}$, $q_1=\sqrt{G}\nu_{1}m_1$ with $\nu_1=0.72$; $m_2=20M_{\odot}$, $q_2=\sqrt{G}\nu_{2}m_2$ with $\nu_2=-0.32$. Magnetic field strength: $\bm{B}=8.0\times 10^{12}\,\mathrm{G}$. Initial semi-major axis: $a=50GM/c^2$ for the uncharged cases and $a=90GM/c^2$ for the charged cases (chosen to enhance radiative effects). Luminosity distance $L=1\,\mathrm{Mpc}$.}
    \label{fig:plus and cross time series}
\end{figure}

Figure~\ref{fig:plus and cross time series} displays the time-domain evolution of cross polarization components. For comparison, we plot the waveforms for systems with charges of the same sign in the magnetic field. In the absence of magnetic fields, orbital motion remains confined to a plane at all PN orders, resulting in vanishing cross polarization when the observation orientation is along $X$-axis (the edge-on case). The coalescence time in Case 2 is substantially shorter than in the other configurations, particularly compared to Case 3. This accelerated inspiral arises because electromagnetic radiation reaction enters at 1.5PN order, exceeding the gravitational contribution by a factor of $\sim v/c$. Conversely, in Case 1, the external magnetic field prolongs the inspiral through the velocity-dependent magnetic Lorentz force, which strengthens as the orbital velocity increases during inspiral. The  secular precession of the angular momentum breaks the planar symmetry, resulting in a rapid growth of the cross polarization in Case 1, consistent with the analysis for Figure~\ref{fig:waveforms comparison}. Most significantly, the cross polarization has the same order of magnitude as those of Case 2 and 3 in the face-on state.

\begin{figure}
    \centering
    \includegraphics[width=0.45\linewidth]{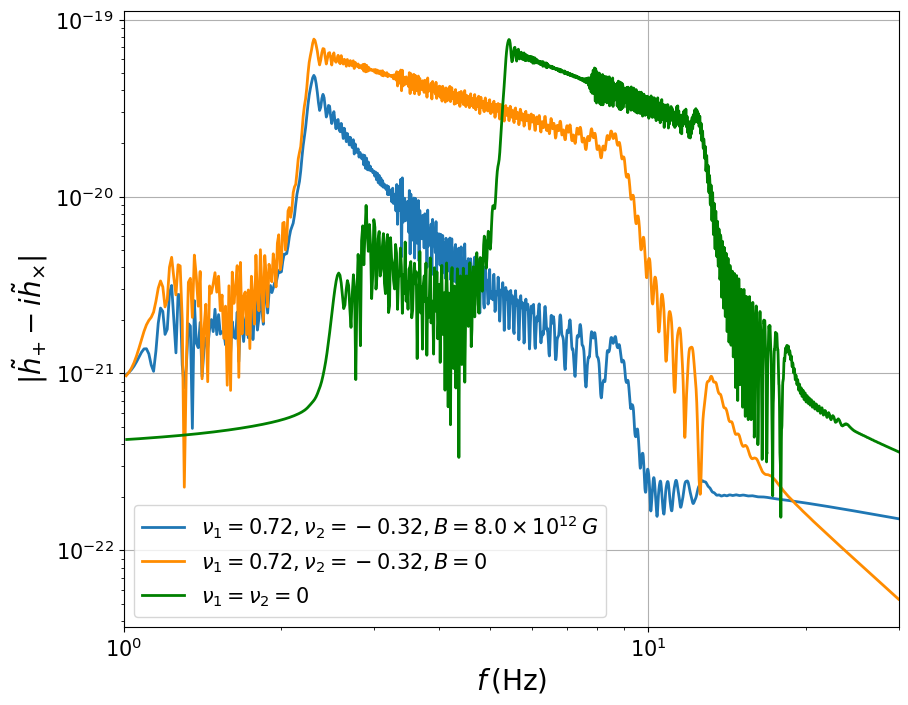}
    \includegraphics[width=0.45\linewidth]{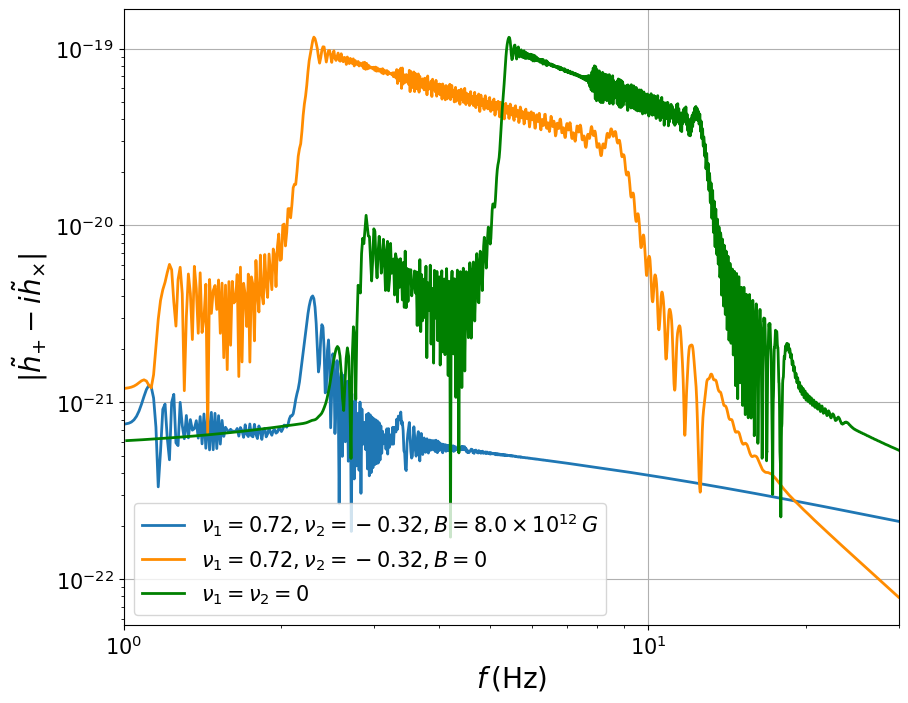}
    \includegraphics[width=0.45\linewidth]{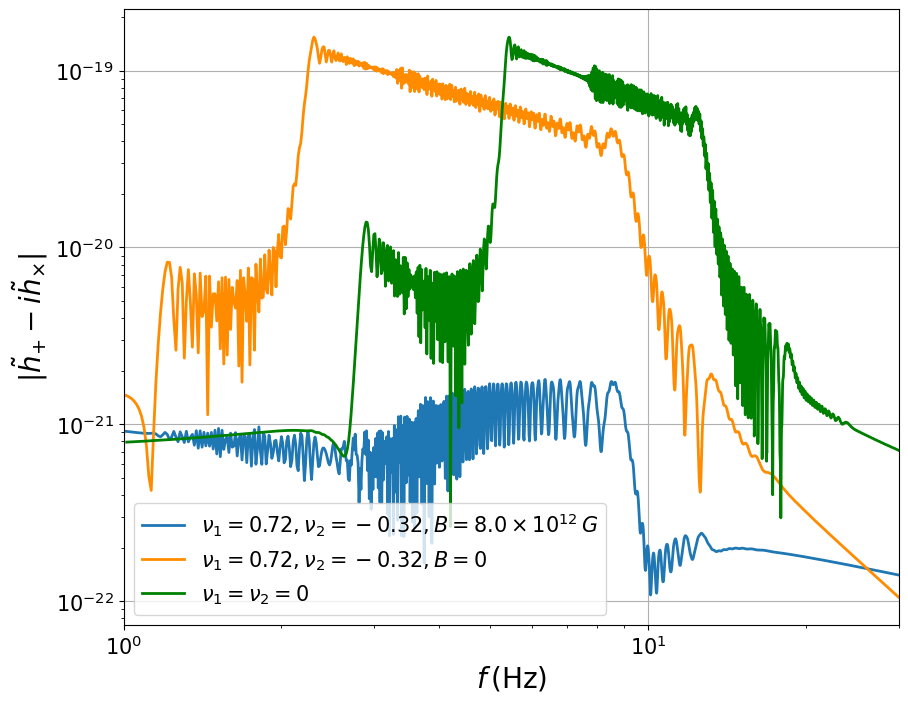}
    \includegraphics[width=0.45\linewidth]{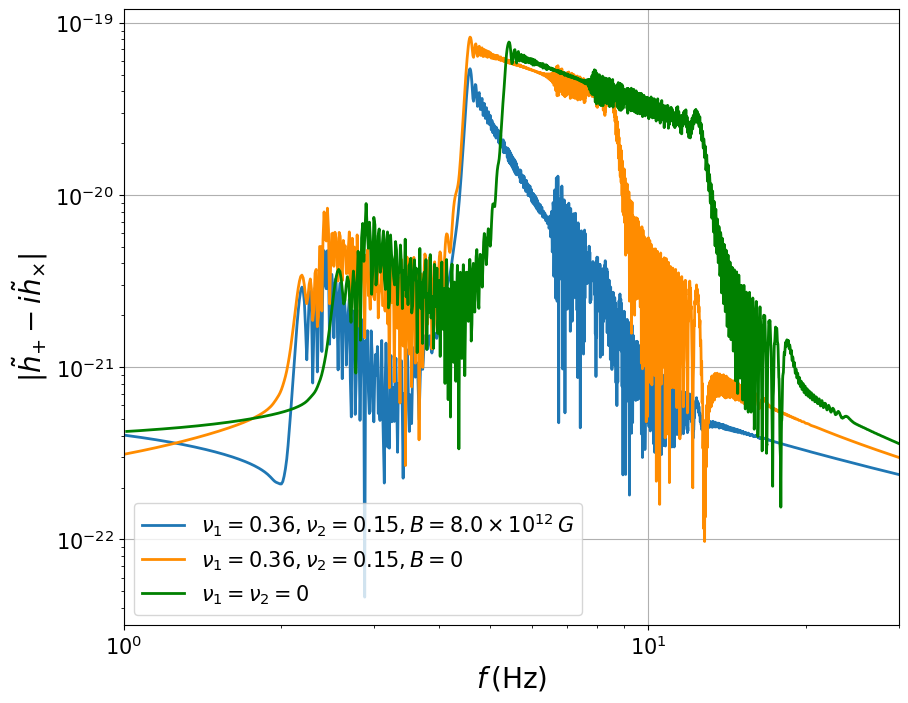}
    \caption{Frequency spectra of $|\tilde{h}_{+}-i\tilde{h}_{\times}|$. Logarithmic scales display the Fourier transforms of the time-domain waveforms from Figure~\ref{fig:plus and cross time series} for all four cases with the same sequence.}
    \label{fig:freq series}
\end{figure}

Figure~\ref{fig:freq series} presents the corresponding frequency spectra obtained through Fourier transformation. The existence of external magnetic field significantly disrupt the power-law scaling of the frequency spectrum, causing the dependence on the direction of observation, which will not occur in Case 2 and 3 \cite{Maggiore:2007ulw,Benavides-Gallego:2022dpn}. Note that the power-law also depends on the charge-to-mass ratio. Such dependence reflects the fact that the magnetic Lorentz force significantly influences the orbital evolution and energy dissipation. The suppression of the magnitude of the frequency spectra in Case 1 comes from the secular drift of CoM which causes an additional evolution of the mass quadrupole moments. The key phenomena provides a distinctive observational signature of magnetically influenced binaries. It is worth noting that the spectra lie in DECIGO's detection band \cite{Kawamura:2020pcg}.

\section{Conclusion and discussion}
\label{sec:conclusion}
In this work, we have developed a comprehensive theoretical framework for charged binary black holes evolving in external magnetic fields, extending the electromagnetic PN formalism to include magnetic field interactions. By incorporating a uniform, stationary magnetic field into the 1PN-expanded Lagrangian and applying the Euler-Lagrange formalism, we derived the complete equations of motion that reveal the intricate interplay between gravitational, electromagnetic, and magnetic field effects. Our numerical investigations demonstrate that external magnetic fields fundamentally alter orbital evolution, most notably breaking the planar symmetry and driving systems into three-dimensional configurations.  Furthermore, the magnetic field introduces substantial contributions to the 1PN corrections in the accelerations, comparable in magnitude to those from charge effects alone, with particularly pronounced enhancements for initially eccentric orbits. Our analysis demonstrates that for certain binary configurations, magnetic field effects imprint distinctive signatures on the gravitational waveforms that fall within the detection capabilities of next-generation interferometers such as DECIGO \cite{Kawamura:2020pcg}.

The implications of our formalism extend to several frontier areas in gravitational physics. Most directly, our results apply to the inspirals of massive, charged black hole binaries, where the 1PN-accurate waveforms we have derived enable robust tests of whether astrophysical black holes carry significant electric charge. The framework also applies to primordial black holes (PBHs)~\cite{Zeldovich:1967lct,Hawking:1971ei,Carr:1974nx}, which some theoretical models suggest may carry substantial charge from early universe processes \cite{Bai:2019zcd, Liu:2020cds}. If charged PBHs constitute dark matter, their passage through galactic magnetic fields would produce the dynamical signatures we have characterized, potentially enabling their identification through combined electromagnetic and gravitational observations.

{Nevertheless, our adoption of a uniform and stationary external magnetic field represents a simplified toy model suitable for preliminary investigations. Realistic astrophysical environments often exhibit dipolar field geometries, which dominate around compact objects throughout the universe. Extending our formalism to such configurations would require incorporating the spatial dependence of $\tilde{F}_{\mu\nu}$. This introduces position-dependent forces and additional gradient terms in the equations of motion, potentially leading to richer orbital dynamics including enhanced precession effects.} A more comprehensive treatment would employ the full PN formalism to capture the coupling between external fields and charges, while simultaneously accounting for field-field interactions within a complete field-theoretical framework. These extensions toward more physically realistic scenarios, including spatially varying fields and self-consistent magnetohydrodynamic effects, constitute our future research agenda.

In conclusion, we have demonstrated that magnetic fields can significantly influence the dynamics and gravitational wave signatures of charged binary black holes, with potentially observable consequences for next-generation detectors. Our results establish a theoretical foundation for incorporating electromagnetic environments into gravitational wave astronomy and highlight the rich phenomenology arising from the interplay of gravity and electromagnetism in strong-field regimes. As gravitational wave observations probe increasingly diverse astrophysical populations, such environmental effects may prove essential for understanding the complete physics of compact binary coalescence.

\begin{acknowledgement}
This work is supported by The National Key R\&D Program of China (Grant No. 2023YFC2206704, No. 2021YFC2203002), NSFC (National Natural Science Foundation of China) No.12473075, No. 12173071, No.12505054, No.12447101, No.12433001 and the Fundamental Research Funds for the Central Universities. This work made use of the High Performance Computing Resource in the Core Facility for Advanced Research Computing at Shanghai Astronomical Observatory.
\end{acknowledgement}

\bibliographystyle{spphys}
\bibliography{Ref}

\end{document}